\documentclass[onecolumn,12pt]{revtex4}

\usepackage{graphicx}
\usepackage{dcolumn}
\usepackage{bm}

\input epsf

\begin{document}

\title{\Large Role of Chameleon Field in presence of Variable Modified Chaplygin gas in Brans-Dicke Theory}

\author{\bf Shuvendu Chakraborty$^1$\footnote{shuvendu.chakraborty@gmail.com} and Ujjal
Debnath$^2$\footnote{ujjaldebnath@yahoo.com ,
ujjal@iucaa.ernet.in}}

\affiliation{$^1$Department of Mathematics, Seacom Engineering College, Howrah-711 302, India.\\
$^2$Department of Mathematics, Bengal Engineering and Science
University, Shibpur, Howrah-711 103, India.}

\date{\today}

\begin{abstract}
In this work, we have considered FRW model of the universe for
Brans-Dicke (BD) theory with BD scalar field as a Chameleon field.
First we have transformed the field equations and conservation
equation from Jordan's frame to Einstein's frame.  We have shown
in presence of variable modified Chaplygin gas, the potential
function $V$ and another analytic function $f$ always increase
with respect to BD-Chameleon scalar field $\phi$.
\end{abstract}

\maketitle

\newpage

\section{\normalsize\bf{Introduction}}
The present accelerated expansion of the universe are strongly
believed under some observations of the luminosity-redshift
relation for type-Ia supernovae [1, 2], measurements of CMBR,
power spectrum of mass perturbation etc. The responsible candidate
for the violation of strong energy condition is known as dark
energy. Another simplest alternative candidate which includes the
scalar field in addition to the tensor field in general relativity
is Brans-Dicke (BD) theory. Recently, enormous interest in BD
theory [3] can be found to the discovery by La et al [4] that the
use of BD theory in place of general relativity can upgrade the
problem of inflationary cosmology. This is possible as it arises
naturally as the low energy limit due to the interaction of the BD
scalar field. The field equations in this case are modified due to
the coupling between matter Lagrangian and scalar field. In BD
theory, there exists one dimensionless parameter $\omega$ and the
effective gravitational constant $G$ which is inversely
proportional to the scalar field $\phi$. Also, BD theory or its
modifications become an important theory to explain the present
cosmic acceleration [5] with some correction of Newtonian
weak-field limit [6]. Einstein gravity can be reconstructed from
it considering the condition $\omega \rightarrow \infty$ [7].
Several authors have done extensive amount of work on BD theory
and generalized BD theory to solve different cosmological problems
of inflation, quintessence etc [8 - 12].\\

There are different dark energy models which are basically
considered as a scalar field, suggest a large correction to the
Newton's law. These are nearly massless scalar field and if exists
on earth, should have been detected in local test of equivalence
principle and as a fifth force. Recently, a different approach is
introduced in general relativity by Khoury and Weltman [13], where
a non-minimal interaction occurs between the quintessence scalar
field and matter sector rather than the geometry and this
interaction is introduced through an interference term in the
action known as Chameleon scalar field. The important feature of
this field is that due to the coupling to matter it obtains an
effective potential i.e., an effective mass depends quite
extremely on the background energy density. Many authors have been
studied on Chameleon field in different aspects of accelerating
universe [14 - 21].\\

A well known candidate for Q-matter is the exotic type of fluid
named Chaplygin gas obeying the EOS $p=-\frac{B}{\rho}$ ($B>0$)
[22], where $p$ and $\rho$ are respectively the pressure and
energy density. A generalized version of this Q-matter having the
form of EOS given by $p=-\frac{B}{\rho^{\alpha}}~(0\leq\alpha\leq
1)$ [23,24]  and recently it was modified to the form
$p=A\rho-\frac{B}{\rho^{\alpha}}$ ($A>0$) [25-27], which is known
as modified Chaplygin Gas. This model can represents the
evaluation of the Universe starting from the radiation era to the
$\Lambda$CDM model. There is one stage, when the pressure vanishes
and the matter content is equivalent to a pure dust. Recently
another modification have been made by Guo and Jhang [28]. They
proposed variable Chaplygin gas model (VCG) where $B$ is a +ve
function of the cosmological scale factor $a$. This $B(a)$ is
related to the scalar potential if we
take the Chaplygin gas Born-Infeld scalar field [29].\\

From these three models (BD theory, Chameleon field and VCG) we
have construct our paper as follows: BD theory or its
modifications have recently described as a driver of the present
cosmic acceleration. It had been shown in [5] that BD theory can
potentially generate sufficient acceleration in the matter
dominated era even without including the exotic Q-field [5]. But
this has problems to establish the "transition" from a decelerated
to an accelerated phase. On the other hand, this transition is
possible when a Chameleon scalar field have a non-minimal coupling
with dark matter [20,21]. Here we consider the BD scalar field as
a Chameleon type scalar field and the matter Lagrangian is coupled
with an arbitrary analytical function of the field. Initially, we
write the field equation in most general scalar tensor theory of
gravity but the field equations become complicated for the
gravitational and scalar fields. So we transfer these equation in
Einstein frame using conformal transformation [30,31], in which
the scalar and tensor degrees of freedom do not mix. Now we
consider the universe filled with variable modified Chaplygin gas
(VMCG) [32] and in general scalar tensor gravity, we graphically
analyze the potential of the BD-Chameleon scalar field  with
respect to the field. We also analyze the analytical function of
the BD-Chameleon scalar field.\\

\section{\normalsize\bf{Basic Equations and Solutions}}
We consider here the Brans-Dicke scalar field $\phi$ act as a
Chameleon field and we consider the action for the chameleon
scalar field $\phi$ in self-interacting Brans-Dicke (BD) theory
[31] as: (choosing $8\pi G_{0}=c=1$)

\begin{equation}
S=\int d^{4} x \sqrt{-g}\left[\phi R- \frac{\omega}{\phi}
{\phi}^{,\alpha} {\phi,}_{\alpha}-V(\phi)+ f(\phi){\cal
L}_{m}\right]
\end{equation}

where $V(\phi)$ is the self-interacting potential for the
BD-Chameleon scalar field $\phi$ and $\omega$ is the BD parameter.
Unlike the usual Einstein-Hilbert action, the matter lagrangian
$L_{m}$ is considered as $f(\phi)L_{m}$, where $f(\phi)$ is some
arbitrary  analytical function of $\phi$. This term makes possible
the non-minimal interaction between the cold
dark matter and BD-Chameleon scalar field.\\

The metric of a homogeneous and isotropic universe in the FRW
model is

\begin{equation} ds^{2}= dt^{2}-a^{2}(t)\left[\frac{dr^{2}}{1-kr^{2}}+
r^{2}(d\theta^{2}+\sin^{2}\theta d\phi^{2}\right]
\end{equation}

where $a(t)$ is the scale factor and $k ~(= ~0,~ +1,~ -1)$ is the
curvature scalar. Now, in  BD  theory, the Einstein's  field
equations  for  the above space-time  symmetry  are

\begin{equation}
H^{2}+\frac{k}{a^{2}}-\frac{\omega}{6}\frac{\dot{\phi}^{2}}{\phi^{2}}+H\frac{\dot{\phi}}{\phi}=\frac{f(\phi)}{3
\phi} \rho +\frac{V(\phi)}{6 \phi}
\end{equation}

\begin{equation}
2\frac{\ddot{a}}{a}+H^{2}+\frac{k}{a^{2}}+\frac{\omega}{2}\frac{\dot{\phi}^{2}}{\phi^{2}}+2H
\frac{\dot{\phi}}{\phi}+\frac{\ddot{\phi}}{\phi}=-\frac{f(\phi)}{
\phi}p +\frac{V(\phi)}{2 \phi}
\end{equation}

and the wave equation for the BD-Chameleon scalar field $\phi$ is

\begin{equation}
\frac{\ddot{\phi}}{\phi}+3H
\frac{\dot{\phi}}{\phi}=\frac{\rho-3p}{(2
\omega+3)\phi}\left(f(\phi)-\frac{1}{2}\phi\frac{df}{d
\phi}\right)+\frac{2}{2
\omega+3}\left(V(\phi)-\frac{1}{2}\phi\frac{dV}{d \phi}\right)
\end{equation}

The energy conservation equation is given by

\begin{equation}
\dot{\rho}+\frac{\dot{a}}{a}(\rho+p)=-\frac{3}{4}(\rho+p)\frac{\dot{f}(\phi)}{f(\phi)}
\end{equation}
Here we consider the variable modified Chaplygin gas with equation
of state [32]

\begin{equation}
p = A\rho- \frac{B}{\rho^{\alpha}},A > 0,B > ,0\leq \alpha \leq 1
\end{equation}

where $B$ is not constant. In the BD theory, the effective
gravitational constant is given by $G = \frac{G_{0}}{\phi}$, which
is indeed not a constant. Now considering the the transformation

\begin{equation}
\bar{g}_{\mu\nu}= \phi g_{\mu\nu}~,
\end{equation}

so that the equations in the Jordan's frame reduce to the
equations in the Einstein's frame. So let us define the new
variables in Einstein's frame as in the following:

\begin{equation}
\psi= \ln\phi, \bar{a}^{2}= \phi a^{2},\rho= \phi\bar{\rho},p=
\phi\bar{p}, f=\phi\bar{f},V=\phi^{2}\bar{V},d\tau=\sqrt{\phi}~dt
\end{equation}

Now the BD field equations (3), (4) and the conservation equation
(6) can be reduced to the Einstein frame as

\begin{equation}
\frac{\bar{a}'^{2}}{\bar{a}^{2}}+\frac{k}{\bar{a}^{2}}=\left(\frac{2
\omega+3
}{12}\right)\psi'^{2}+\frac{\bar{f}(\psi)\bar{\rho}}{3}+\frac{\bar{V}(\psi)}{6}
\end{equation}
\begin{equation}
2\frac{\bar{a}''}{\bar{a}}+\frac{\bar{a}'^{2}}{\bar{a}^{2}}+\frac{k}{\bar{a}^{2}}=-\left(\frac{2
\omega+3
}{4}\right)\psi'^{2}-\bar{p}\bar{f}(\psi)+\frac{\bar{V}(\psi)}{2}
\end{equation}
and
\begin{equation}
\bar{\rho}'+3\frac{\bar{a}'}{\bar{a}}(\bar{\rho}+\bar{p})=-\frac{1}{4}(\bar{\rho}-3\bar{p})\psi'-\frac{3}{4}(\bar{\rho}+\bar{p})\frac{\bar{f}'}{\bar{f}}
\end{equation}

where, $'$ represents the derivative w.r.t. new time variable
$\tau$. Also using the above transformation and considering the
variable term $B= B_{0}\phi^{\alpha+1}$, the EOS of the variable
modified Chaplygin gas (7) becomes
\begin{equation}
\bar{p}=A \bar{\rho}-\frac{B_{0}}{\bar{\rho}^{\alpha}}
\end{equation}

where $B_0(>0)$ is a constant. This EOS represents only the
modified Chaplygin gas in the Einstein's frame. From the equations
(12) and (13) we can construct a linear differential equation in
$z$ (where, $\bar{\rho}=z^{\frac{1}{\alpha+1}}$) as

\begin{equation}
\frac{dz}{d\tau}+z(\alpha+1)\left(
3(1+A)\frac{\bar{a}'}{\bar{a}}+\frac{3}{4}(1+A)\frac{\bar{f}'}{\bar{f}}+\frac{1-3A}{4}\psi'\right)=(\alpha+1)
\left(3B_{0}\frac{\bar{a}'}{\bar{a}}+\frac{3}{4}B_{0}\frac{\bar{f}'}{\bar{f}}-\frac{3}{4}B_{0}\psi'\right)
\end{equation}

For simplicity we take only the radiation case $A=1/3$ and in this
case the integrating factor (I.F.) becomes
$(\bar{a}^{4}\bar{f})^{\alpha+1}$. We also assume
$\bar{a}^{4}\bar{f}=\phi^{n}$, where $n$ is a positive constant.
After solving the above the differential equation, we get the
solution of $\bar{\rho}$ as

\begin{equation}
\bar{\rho}=\left[\frac{3}{4}B_{0}\left(
1-\frac{\alpha+1}{n(\alpha+1)+1}\psi+C
\psi^{-n(\alpha+1)}\right)\right]^{\frac{1}{\alpha+1}}
\end{equation}
where $C$ is an integrating constant. Now let the transformed
Hubble parameter is defined by $\frac{\bar{a}'}{\bar{a}}=\bar{H}$
then $\bar{H}'=\bar{a}\bar{H}\frac{d\bar{H}}{d\bar{a}}$. Now we
take a new assumption $\psi=\psi_{0}\bar{a}^{m}$ where $m$ is a
positive constant. Using this new assumption the solution of
$\bar{\rho}$ becomes
\begin{equation}
\bar{\rho}=\left[\frac{3}{4}B_{0}\left(
1-\frac{\alpha+1}{n(\alpha+1)+1}\psi_{0}\bar{a}^{m}+C
(\psi_{0}\bar{a}^{m})^{-n(\alpha+1)}\right)\right]^{\frac{1}{\alpha+1}}
\end{equation}
Now putting $\bar{H}^{2}=\bar{Y}$ and considering the previous all
assumptions, we have from the equations (10) and (11) a
differential equation in $\bar{Y}$ as

\begin{equation}
\frac{d\bar{Y}}{d\bar{a}}+\frac{2
\omega+3}{2}\psi_{0}^{2}m^{2}\bar{a}^{2m-1}\bar{Y}=-(\bar{\rho}+\bar{p})\psi_{0}^{n}\bar{a}^{m
n-5}+\frac{2 k}{\bar{a}^{3}}
\end{equation}

From conformal transformation we have $\bar{a}=\left(\frac{\ln
\phi }{\psi_{0}}\right)^{m}$ which implies
$\frac{d\bar{a}}{d\phi}=\frac{(\ln
\phi)^{\frac{1}{m}-1}\psi_{0}^{-\frac{1}{m}}}{m \phi}$. Now using
equation (13) and (16), we have a transformed form of the equation
(17) given by

 \begin{eqnarray*}
\frac{d\bar{Y}}{d \phi}+\frac{2 \omega+3}{2 \phi}m \ln \phi \
\bar{Y}=-\left[ \frac{4}{3}\left\{ \frac{3
B_{0}}{4}\left(1-\frac{\alpha+1}{n(\alpha+1)}\ln \phi+C(\ln
\phi)^{-n(\alpha+1)} \right)\right\}^{\frac{1}{\alpha+1}}\right.
\end{eqnarray*}

\begin{equation}
\left.-B_{0}\left\{ \frac{3
B_{0}}{4}\left(1-\frac{\alpha+1}{n(\alpha+1)}\ln \phi+C(\ln
\phi)^{-n(\alpha+1)}
\right)\right\}^{-\frac{\alpha}{\alpha+1}}\right]\frac{\psi_{0}^{\frac{4}{m}}(\ln
\psi )^{\frac{n-4}{m-1}}}{m \phi}+\frac{2k}{m \phi}(\ln
\phi)^{-\frac{2}{m-1}}\psi_{0}^{\frac{2}{m}}
 \end{equation}

Using all the substitutions and the equation (9),(10) we get the
expression for the potential $V$ (after simplification) as

\begin{eqnarray*}
V=\phi^{2}\left[\left(6-\frac{2 \omega+3}{2}m^{2}(\ln \phi)^{2}
\right)\bar{Y}+6 k \left(\frac{\ln
\phi}{\phi_{0}}\right)^{-\frac{2}{m}}
\right.~~~~~~~~~~~~~~~~~~~~~~~~~~~~~~
\end{eqnarray*}
\begin{equation}
\left. -2(\ln \phi)^{\frac{n-4}{m}}\psi_{0}^{\frac{4}{m}} \left\{
\frac{3 B_{0}}{4}\left(1-\frac{\alpha+1}{n(\alpha+1)}\ln
\phi+C(\ln \phi)^{-n(\alpha+1)}
\right)\right\}^{\frac{1}{\alpha+1}} \right]
\end{equation}
where $\bar{Y}$ is the solution of the equation (18).\\

The analytical function $f$ in terms of $\phi$ is given by
\begin{equation}
f=\phi \psi_{0}^{n}\left(\frac{\ln
\phi}{\phi_{0}}\right)^{\frac{m n -4}{m n}}
\end{equation}
We now graphically represent $V$ against $\phi$ and $f$ against
$\phi$. From the figure 1 we see that considering BD scalar field
as Chameleon, the potential $V(\phi)$ is sharply increasing with
respect to the field and from figure 2 we see that the analytical
function $f(\phi)$ is also increasing as field increases for some
particular values of the parameters i.e., for $\alpha = 0.5, B_{0}
= 4, C = 1, m = 4, n = 2, \psi_{0}=1, \omega=-1/3$.\\

\begin{figure}
\includegraphics[scale=0.7]{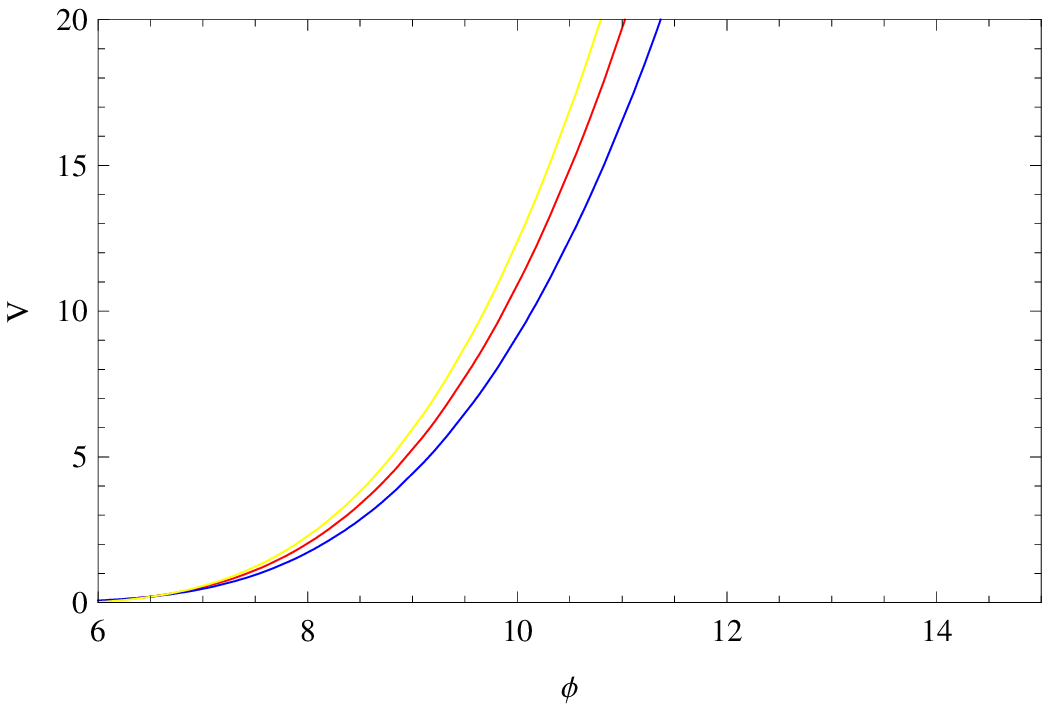}~~~~~~~~
\includegraphics[scale=0.7]{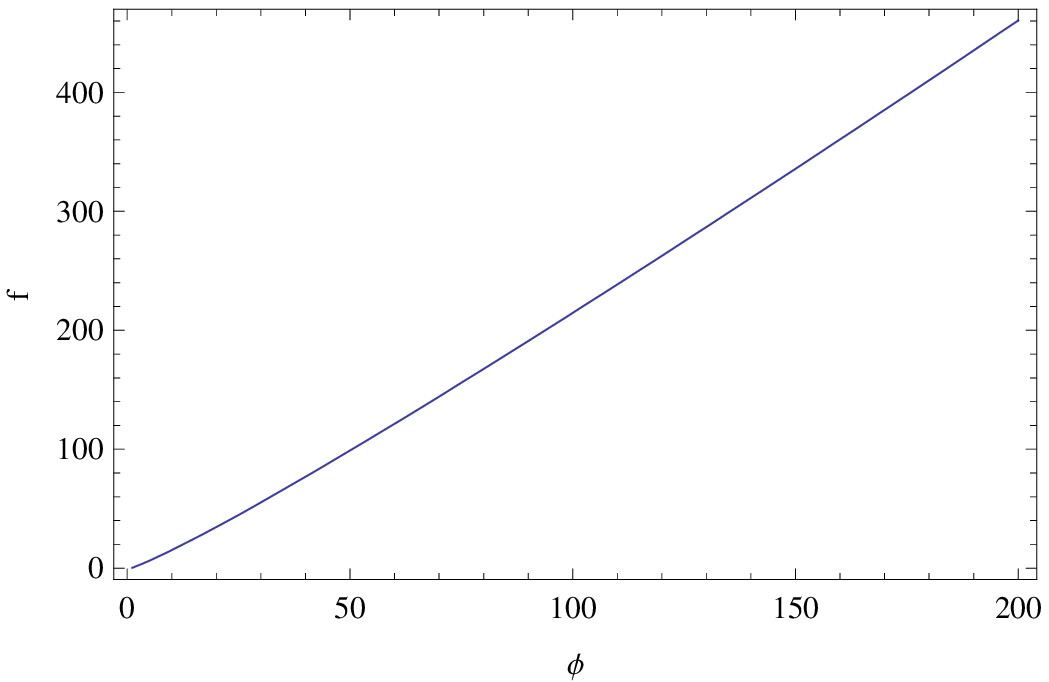}\\
\vspace{1mm} ~Fig.1~~~~~~~~~~~~~~~~~~~~~~~~~~~~~~~~~~~~~~~~~~~~~~~~~~~~~~~~Fig.2\\

\vspace{6mm} Fig. 1 shows the variation of $V$ against $\phi$ for
flat (red curve), closed (blue curve), open (yellow curve)
universe respectively and fig. 2 shows the variation of $f$
against $\phi$ for $\alpha = 0.5, B_{0} = 4, C = 1, m = 4, n = 2,
\psi_{0}=1, \omega=-1/3$.

 \vspace{6mm}
\end{figure}

\section{\normalsize\bf{Discussions}}

Here we have considered initially general scalar tensor
Brans-Dicke theory of gravity having BD scalar field as Chameleon
and then using conformal transformation we rewrite the field
equations in Einstein's frame. Now considering the universe filled
with variable modified Chaplygin gas and using some particular
assumptions, we solve the density parameter for the radiation era.
Now we graphically analyzed the potential of BD-Chameleon scalar
field with respect to the field. We see from figure 1 that
considering BD scalar field as a Chameleon and in presence of VMCG
the potential is increasing with the increase of the field. We
also see from the figure 2 that the analytical function of the
field which is coupled with the matter Lagrangian in the action is
also increasing with the increase of the BD-Chameleon scalar field.\\

{\bf Acknowledgement}:\\

The authors are thankful to IUCAA, Pune, for their warm
hospitality and excellent research facilities where part of the
work has been done during a visit.\\

{\bf References:}\\
\\
$[1]$ S. J. Perlmutter et al, {\it  Astrophys. J.} {\bf 517} 565 (1999).\\\
 $[2]$ A. G. Rieses et al, {\it Astron. J.} {\bf
116} 1009 (1998); P. M. Garnavich et al,
{\it Astrophys. J.} {\bf 509} 74 (1998); G. Efstathiou et al, astro-ph/9812226.\\\
$[3]$ C. Brans and R. H. Dicke, {\it Phys. Rev.} {\bf 124} 925
(1961).\\
$[4]$ D. A. La and P. J. Steinhardt, {\it Phys. Rev. Lett.} {\bf
62} 376 (1989).\\
$[5]$ N. Banerjee and D. Pavon, {\it Phys. Rev. D} {\bf 63} 043504 (2001).\\
$[6]$ C. Will, {\it Theory and Experiments in Gravitational
Physics} (Cambridge, Cambridge University Press) (1993).\\
$[7]$ B. K. Sahoo and L. P. Singh, {\it Mod. Phys. Lett. A} {\bf 18} 2725 (2003).\\
$[8]$ K. Nordtvedt, {\it Astrophys. J} {\bf 161} 1059 (1970); P.
G. Bergmann, {\it Int. J. Phys.} {\bf 1} 25 (1968); R. V. Wagoner,
{\it Phys. Rev. D} {\bf 1} 3209 (1970); T. Damour and K.
Nordtvedt, {\it Phys. Rev. Lett.} {\bf 70} 2217 (1993); {\it Phys. Rev. D} {\bf 48} 3436 (1993).\\
$[9]$ P. G. Bergmann, {\it Int. J. Theor. Phys.} {\bf 1} 25
(1968); A. Sheykhi and M. Jamil, arXiv:1010.0385v3; D. J. Holden
and D. Wands, {\it Class. Quantum Grav.} {\bf 15} 3271 (1998); R. V. Wagoner,
{\it Phys. Rev. D} {\bf 1} 3209 (1970); S. J. Kolitch and D. M. Eardley,
{\it Annals of Physics} {\bf 241} 128 (1995) .\\
$[10]$ J. D. Barrow and K. Maeda, {\it Nucl. Phys. B} {\bf 341} 294 (1990).\\
$[11]$ C. Santos and R. Gregory, {\it Ann. Phys., (NY)} {\bf 258} 111 (1997).\\
$[12]$ O. Bertolami and P. J. Martins, {\it Phys. Rev. D} {\bf 61} 064007 (2000).\\
$[13]$ J. Khoury and A. Weltman, \textit{Phys. Rev. Lett.}, \textbf{93} 171104 (2004).\\\
$[14]$ A. -C. Davis, C. A. O. Schelpe and D. J. Shaw, \textit{Phys. Rev. D} \textbf{80} 064016 (2009).\\\
$[15]$ Y. Ito and S. Nojiri, \textit{Phys. Rev. D} \textbf{79} 103008 (2009).\\\
$[16]$ H. Sanctuary and R. Sturani, arXiv:0809.3156v1[gr-qc].\\\
$[17]$ T. Tamaki and S. Tsujikawa,  \textit{Phys. Rev. D} \textbf{78} 084028 (2008).\\\
$[18]$ P. Brax, C. van de Bruck, A. -C. Davis and D. J. Shaw,  \textit{Phys. Rev. D} \textbf{78} 104021 (2008).\\\
$[19]$ A. Weltman,  arXiv:0805.3461v1[hep-th].\\\
$[20]$ N. Banerjee, S. Das and K. Ganguly,  \textit{Pramana} \textbf{74} L481 (2010).\\\
$[21]$ S. Chakraborty and U. Debnath, \textit{Int. J. Mod. Phys. A}   \textbf{25} 24 (2010).\\\
$[22]$ A. Kamenshchik, U. Moschella and V. Pasquier, \textit{Phys.
Lett. B} \textbf{511} 265 (2001).\\
$[23]$ V. Gorini, A. Kamenshchik and U. Moschella, {\it Phys. Rev.
D} {\bf 67} 063509 (2003).\\
$[24]$ M. C. Bento, O. Bertolami and A. A. Sen, \textit{Phys. Rev.
D} \textbf{66} 043507 (2002).\\
$[25]$ H. B. Benaoum, hep-th/0205140.\\
$[26]$ U. Debnath, A. Banerjee and S. Chakraborty, \textit{Class.
Quantum Grav.} \textbf{21} 5609 (2004).\\
$[27]$ S. Chakraborty and U. Debnath, \textit{Astrophys. Space
Sci.} \textbf{321} 53 (2009).\\
$[28]$ Z. K. Guo and Y. Z. Zhang,
\textit{Phys. Lett. B} \textbf{645} 326 (2007).\\
$[29]$ M. C. Bento, O. Bertolami and A. A. Sen, \textit{Phys.
Lett. B} \textbf{575} 172 (2003).\\
$[30]$ V. B. Bezerra et al, \textit{Braz. J. Phys.} \textbf{34}
2A (2004).\\
$[31]$ S. Das and N. Banerjee, \textit{Gen. Rel. Grav.}
\textbf{38} 785 (2006).\\
$[32]$ U. Debnath, {\it Astrophys. Space Sci.} {\bf 312} 295
(2007).\\

\end{document}